\documentclass[letterpaper, 10 pt, conference]{ieeeconf}  
\IEEEoverridecommandlockouts                              
\let\labelindent\relax
\usepackage{enumitem}
\setlist{nosep, leftmargin=14pt}
\usepackage{graphicx}
\usepackage{hyperref}       
\usepackage{url}            
\usepackage{booktabs}       
\usepackage{amsfonts}       
\usepackage{latexsym}
\usepackage{nicefrac}       
\usepackage{microtype}      
\usepackage{xcolor}
\usepackage{multirow}       
\usepackage{threeparttable} 
\usepackage{amsmath}        
\DeclareMathOperator*{\argmin}{argmin}

\title{\LARGE \bf
Mutual Information Neural Estimation for \\ Unsupervised Multi-Modal Registration of Brain Images
}
\author{Gerard Snaauw$^{1}$, Michele Sasdelli$^{1}$, Gabriel Maicas$^{1}$, Stephan Lau$^{1,2}$, \\ Johan Verjans$^{1,2}$, Mark Jenkinson$^{1,2,*}$, and Gustavo Carneiro$^{1}$%
\thanks{$^1$Australian Institute for Machine Learning (AIML), University of Adelaide, Adelaide, Australia, $^2$South Australian Health and Medical Research Institute (SAHMRI), Adelaide, Australia, *Correspondence may be directed to Prof. Mark Jenkinson: {\tt\small mark.jenkinson@adelaide.edu.au}}%
}

\begin{document}

\onecolumn
{
\vspace*{6cm}
\LARGE
\noindent G. Snaauw et al., "Mutual Information Neural Estimation for Unsupervised Multi-Modal Registration of Brain Images," 2022 44th Annual International Conference of the IEEE Engineering in Medicine \& Biology Society (EMBC), 2022, pp. 3510-3513, doi: 10.1109/EMBC48229.2022.9871220.\\
\vspace*{1cm}

\noindent © 2022 IEEE. Personal use of this material is permitted. Permission from IEEE must be obtained for all other uses, in any current or future media, including reprinting/republishing this material for advertising or promotional purposes, creating new collective works, for resale or redistribution to servers or lists, or reuse of any copyrighted component of this work in other works.

}
\twocolumn
\newpage

\maketitle
\thispagestyle{empty}
\pagestyle{empty}

\begin{abstract}
Many applications in image-guided surgery and therapy require fast and reliable non-linear, multi-modal image registration. Recently proposed unsupervised deep learning-based registration methods have demonstrated superior performance compared to iterative methods in just a fraction of the time. Most of the learning-based methods have focused on mono-modal image registration. The extension to multi-modal registration depends on the use of an appropriate similarity function, such as the mutual information (MI). We propose guiding the training of a deep learning-based registration method with MI estimation between an image-pair in an end-to-end trainable network. Our results show that a small, 2-layer network produces competitive results in both mono- and multi-modal registration, with sub-second run-times. Comparisons to both iterative and deep learning-based methods show that our MI-based method produces topologically and qualitatively superior results with an extremely low rate of non-diffeomorphic transformations. Real-time clinical application will benefit from a better visual matching of anatomical structures and less registration failures/outliers. 
\end{abstract}

\section{Introduction}
\label{sec:intro}

Multi-modal registration has many use cases, such as in image-guided therapy where pre-therapeutic imaging is aligned with live data to guide treatment (e.g., tumor localization in surgery or radiotherapy). The large computational run-time of conventional iterative methods mean that registration methods in these applications are limited to inaccurate linear transforms~\cite{Sotiras2013}. Recent advances in unsupervised deep learning-based registration have demonstrated sub-second run-times in accurate mono-modal deformable registration~\cite{Balakrishnan2019}. A range of auxiliary approaches have since been investigated to improve registration performance, specifically segmentation maps, multi-resolution, and locally adaptive regularisation~\cite{Balakrishnan2019,Dalca2019,Nan2020,Zhu2021, Qiu2021}. The extension of deep learning methods to multi-modal registration depends on an appropriate similarity function, such as mutual information (MI)~\cite{Wells1996}, which is one of the most studied and successful similarity metrics in multi-modal iterative registration approaches~\cite{Sotiras2013}. Application of MI in a learning-based setting is not trivial due to the non-differential quantization step (i.e., binning of intensity values). One solution is to use a continuous quantization function~\cite{Qiu2021}.

In this paper, we propose estimating MI using a 2-layer convolutional neural network (CNN) to extend deep learning-based frameworks to multi-modal registration. More specifically, we estimate MI with mutual information neural estimation (MINE) that eliminates the cumbersome discretization of the density estimation step and allows end-to-end training as it is fully differentiable~\cite{Belghazi2018}.

\section{Methods} 
\label{sec:methods}
\subsection{Registration framework} 
Image registration estimates a transform $\phi_{\mathbf{z}}:\Omega \rightarrow \mathbb{R}^3$ (parameterized by $\mathbf{z}$) that aligns a moving image $\mathbf{m}:\Omega \rightarrow \mathbb{R}$ to a fixed image $\mathbf{f}:\Omega \rightarrow \mathbb{R}$, such that $\mathbf{m}(\phi_{\mathbf{z}}(\omega))$ becomes similar to $\mathbf{f}(\omega)$,
where $\omega$ are points on the image lattice $\Omega$.
For brevity we drop $\omega$ and write transformation as $\mathbf{m}\circ\phi_{\mathbf{z}}$.

\begin{figure*}[ht]
    \centering
    \includegraphics[width=12cm]{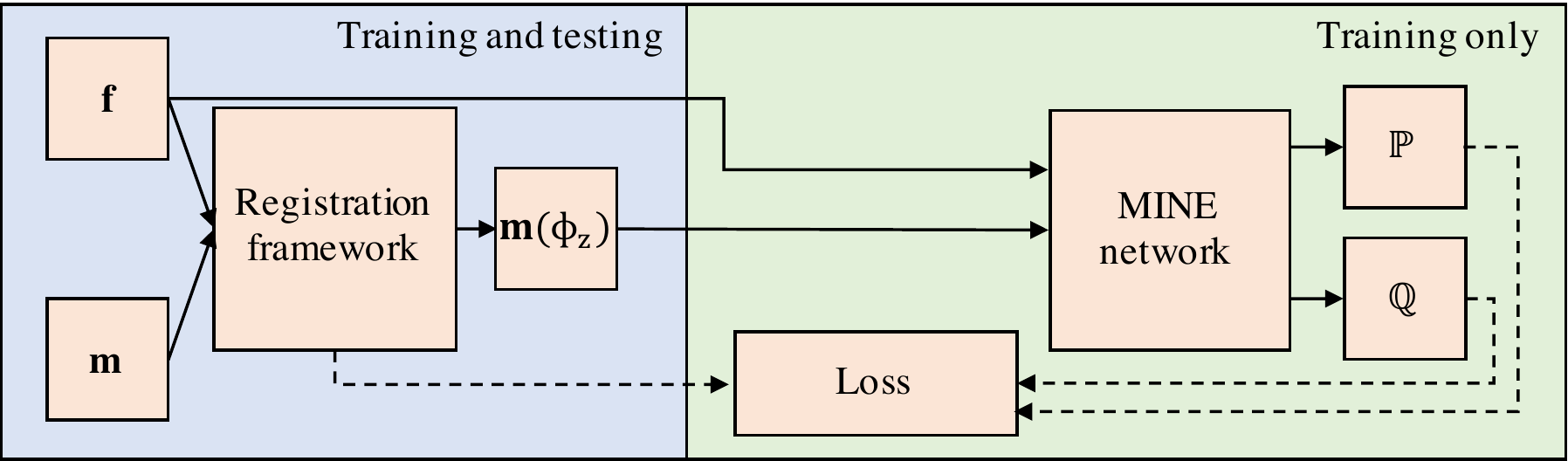}
    \caption{Overview of the unsupervised registration approach using MINE. Left illustrates a generic unsupervised registration framework that produces the transform to align an image pair. Right shows our proposed method for training the framework by estimating MI with MINE. The MINE network forms a single end-to-end differentiable network with the registration framework during training, and can be removed hereafter.}
    \label{fig:MINE}
\end{figure*}

For our registration framework we use the diffeomorphic variant of VoxelMorph (VM)~\cite{Dalca2019}, see Fig.~\ref{fig:MINE}, but our method is not specific to this framework. VM takes the input images $\mathbf{f},\mathbf{m}$ and passes them through a Unet-like structure~\cite{Ronneberger2015} to output a posterior distribution from which a sample is taken that is a stationary velocity field~\cite{Dalca2019}. This is then integrated to form a diffeomorphic transform $\phi_{\mathbf{z}}$ using a scaling and squaring approach~\cite{Arsigny2006}, and applied to the moving image. VM applies the mean square error (MSE), a mono-modal similarity metric, between $\mathbf{m}\circ\phi_{\mathbf{z}}$ and $\mathbf{f}$ and a regularization function that contains one hyper-parameter $\lambda$ to control the scale of the velocity field.

\subsection{Novel image matching loss function} 
We replace the image matching term in the loss function of VM~\cite{Dalca2019} with a loss function based on MI that has been demonstrated to work effectively for multi-modal problems in traditional (non-machine-learning) registration methods~\cite{Wells1996}. The design of our image matching loss function is based on MINE~\cite{Belghazi2018} and replaces the cumbersome approximation of MI through a binning approach by training of a network that estimates MI and works on multi-dimensional continuous-valued data. 

The MI-based registration of $\mathbf{m}$ to $\mathbf{f}$ is defined by
\begin{equation} 
    I(\mathbf{f},\mathbf{m}) = \int_{\Omega} 
        p(\mathbf{f}, \mathbf{m}\circ\phi_{\mathbf{z}}) \: 
        \log\left( \frac{
            p(\mathbf{f}, \mathbf{m}\circ\phi_{\mathbf{z}} )}
            {p(\mathbf{f}) p(\mathbf{m}\circ\phi_{\mathbf{z}})} 
        \right)d\omega,
    \label{eq:mi:distributions} 
\end{equation}
where $p(\mathbf{f}, \mathbf{m}\circ\phi_{\mathbf{z}})$ measures the joint probability between the gray values of the fixed and moving images, and
$p(\mathbf{f})$ and  $p(\mathbf{m}\circ\phi_{\mathbf{z}})$ are the marginal probabilities of $\mathbf{f}$ and $\mathbf{m}\circ\phi_{\mathbf{z}}$ 
-- note that the MI in~\eqref{eq:mi:distributions} is equivalent to the Kullback-Leibler (KL) divergence between the joint-probability density $\mathbb{P}=p(\mathbf{f},\mathbf{m} \circ \phi_{\mathbf{z}})$ 
and product of marginals $\mathbb{Q}=p(\mathbf{f}) p(\mathbf{m} \circ \phi_{\mathbf{z}})$, 
defined by $D_{KL} \left(\ \mathbb{P} \ || \ \mathbb{Q}\ \right)$.  
We extend MINE~\cite{Belghazi2018} by training the image matching network with a lower-bound estimate on the MI in~\eqref{eq:mi:distributions} using the following inequality~\cite{Donsker1983}:
\begin{equation} 
    D_{KL} \left(\:\mathbb{P}\:||\:\mathbb{Q}\:\right) \geq  \sup_{F_{\theta} \in \mathcal{F}}  \mathbb{E}_\mathbb{P}\left[F_{\theta}(\mathbf{f},\mathbf{\hat{m}}) \right] - 
     \log \left( \mathbb{E}_\mathbb{Q}\left[e^{F_{\theta}(\mathbf{f},\mathbf{\hat{m}}) }\right] \right)
    \label{eq:mi:lowerbound}
\end{equation}
where ${F}_{\theta}:\Omega \to \mathbb{R}$ (with ${F}_{\theta} \in \mathcal{F}$) represents any class of functions (parameterized by $\theta$) that satisfies the integrability constraints~\cite{Donsker1983}, 
and $\mathbf{\hat{m}} = \mathbf{m} \circ \phi_{\mathbf{s}} \circ \phi_{\mathbf{z}}$ with
$\phi_{\mathbf{s}}(\omega) = \omega$ when the expectation is over $\mathbb P$ (i.e., $\phi_{\mathbf{s}}$ is the identify function, so $\mathbf{f}$ and $\mathbf{m}$ match according to $\phi_{\mathbf{z}}$), and $\phi_{\mathbf{s}}(\omega) \sim U_{\Omega}(\omega)$ (with $U_{\Omega}(\omega)$ being a flat distribution over $\Omega$, such as a random permutation of locations) when the expectation is over $\mathbb Q$ -- this means that for the product of marginals $\mathbb Q$, $\mathbf{f}$ and $\mathbf{m}$ are unlikely to match according to $\phi_{\mathbf{z}}$. 

The learning function that we use to optimise our model is the following:
\begin{equation} \label{eq:loss}
    \begin{split}
        \theta^*,\psi^* = \argmin_{\theta,\psi} 
        & - \alpha\left(\mathbb{E}_\mathbb{P}\left[F_{\theta}(\mathbf{f},\mathbf{\hat{m}}) \right] - 
         \log \left( \mathbb{E}_\mathbb{Q}\left[e^{F_{\theta}(\mathbf{f},\mathbf{\hat{m}})}\right] \right)\right) \\
        & + R(\phi_{\mathbf{z}},\lambda),
    \end{split}
\end{equation}
\noindent where $R(.)$ is the regularization function to approximate the variational posterior $q_{\psi}(\mathbf{z}|\mathbf{f},\mathbf{m})$ and the  prior $p(\mathbf{z})$~\cite{Dalca2019}, $\phi_{\mathbf{z}}$ is the diffeomorphic transform computed from the network (see left hand side of Fig.~\ref{fig:MINE}), and $\phi_{\mathbf{s}}$ is defined in Eq.~\eqref{eq:mi:lowerbound}.

In~\eqref{eq:loss}$, F_{\theta}$ is modeled as a simple 2-layer CNN with 1x1x1 filters to preserve the independent voxel assumption in MI. The first layer combines $L$ linear transformed versions of the input to create $L$ features per voxel-pair. Then, a second layer maps these features back to a single value per voxel that are used to generate representations for $\mathbb{P}$ and $\mathbb{Q}$. During inference, we are given two test images $\tilde{\mathbf{m}},\tilde{\mathbf{f}}$, and the registration is obtained by the diffeomorphic transform $\phi_{\mathbf{z}}$.

For the function $U_{\Omega}(\omega)$ in~\eqref{eq:mi:lowerbound}, we evaluate global and local shuffling strategies referred to as MINE$_{global}$ and MINE$_{local}$. In MINE$_{global}$, $\phi_{s}(\omega) \sim {U}_{\Omega}(\Omega)$, which means that it randomly shuffles all voxel coordinates (using a uniform distribution over the whole input lattice $\Omega$) to estimate the distribution over voxel pairs for unaligned images. However, since a large fraction of voxels in our data belong to background, we also evaluate a local shuffling strategy to estimate the distribution of voxel pairs in their expected neighborhood, as would be expected after sensible pre-processing and initialisation. This is denoted by MINE$_{local}$, $\phi_{s}(\omega) \sim {U}_{\Omega}([\omega - N,\omega + N])$. MINE$_{local}$ returns a new input address that is within $2\times N$ (in each direction) of the original address $\omega$.

\subsection{Data and preprocessing}
This research study was conducted retrospectively using human subject data made available in open access~\cite{Marcus2010,Petersen2010}. All ethical standards defined in the data use agreement have been implemented. For our experiments we use the T1-weighted scans from the OASIS-2 dataset~\cite{Marcus2010}, and a subset of the ADNI dataset~\cite{Petersen2010}, where we included all subjects that had a T1-weighted and FLAIR scan with high spatial resolution acquired in the initial session. Preprocessing was performed using FreeSurfer~\cite{Fischl2012} on all T1-weighted scans and consisted of resampling to 1~mm isotropic voxels, affine spatial normalization, brain extraction, and automated generation of segmentation maps. Since both modalities of the ADNI dataset are acquired in the same session we assume that they are well aligned and apply the preprocessing parameters (including segmentation maps) found for T1 to the FLAIR volumes. Anatomical regions were excluded if they had less than 100 voxels in any of the subjects, resulting in 30 regions that will be used for evaluation (see later). Volumes are cropped to $160\times192\times224$ voxels and, after removal of subjects that failed pre-processing, split into training:validation:test sets with 230:24:118 subjects for OASIS and 293:24:118 for ADNI.

\subsection{Training process}
We evaluate our method on two tasks. First, we replicate the setup from~\cite{Dalca2019} and evaluate mono-modal registration to an atlas. Second, we evaluate the performance of our method on multi-modal inter-patient registration by aligning T1-weighted and FLAIR volumes, which we regard as an example test for the non-linear registration required in image-guided procedures. For the training of the MINE-network $F_{\theta(.)}$ in~\eqref{eq:loss}, we set the feature vector length $L=30$, where the neighbouring region for MINE$_{local}$ is set at $N=8$ voxels. Evaluation of these values was limited by available memory and prolonged training and chosen on the basis of capturing a reasonable level of anatomical context (for $N$) and achieving fast and stable convergence (for $L$). In~\eqref{eq:loss}, we set the regularization parameter $\lambda=10$ for all experiments and training was performed using the ADAM optimizer with $\beta_{1}=0.99$, $\beta_{2}=0.999$ and learning rate $10^{-4}$ for 1500 epochs with a batch size of 1 and Kaiming initialization.

\subsection{Comparison with alternative methods} 
The baseline methods that we include are ANTs~\cite{Avants2008} and the diffeomorphic variant of VoxelMorph~\cite{Dalca2019}. ANTs registration uses the default settings that combines SyN~\cite{Avants2008} with cross correlation (CC) and MI to produce the method's optimal results for both registration tasks. The training of VM is performed using local normalized cross-correlation (VM-CC) with neighborhood $9^3$~\cite{Balakrishnan2019}. We use CC rather than MSE as CC is also capable of multi-modal registration.

Evaluation is separated into a quantitative and qualitative part. The quantitative part evaluates 
$1)$ Dice similarity coefficient (DSC) as a measure of registration accuracy, 
$2)$ incidence of non-positive determinants for Jacobian values $|J_\phi(\omega)|\leq0$ as a measure of the diffeomorphic performance, and 
$3)$ run-time at test-time to access feasibility for 
image guided therapies. The qualitative part compares visual registration results and shows the response for the joint distribution $\mathbb P$ in~\eqref{eq:mi:lowerbound}.

\section{Results}

\begin{figure}[bt]
    \begin{center}
        \includegraphics[trim= 33 2 15 8, clip, width=8.6cm]{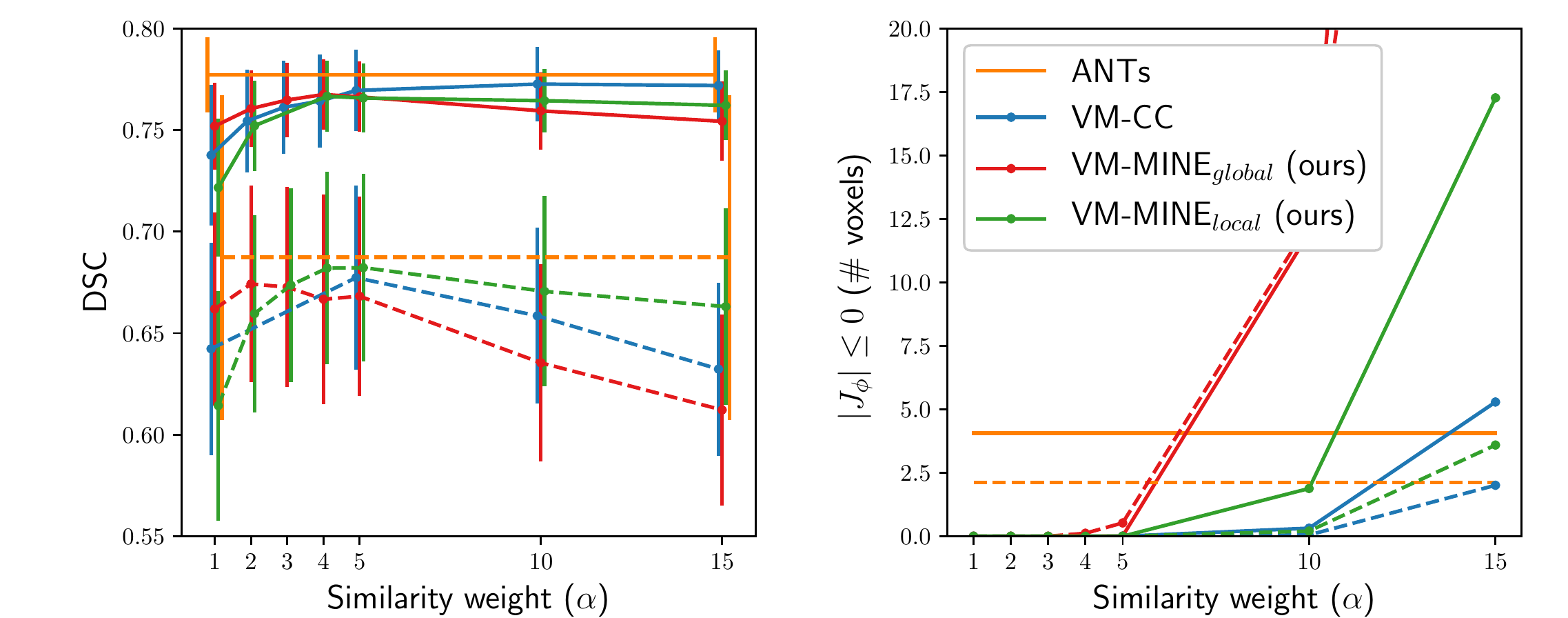}
    \end{center}
    \caption{Performance for different values of similarity weight $\alpha$~\eqref{eq:loss} for mono-modal registration (solid curves) and multi-modal registration (dashed curves). Left: average Dice score. Right: number of non-diffeomorphic voxels.}
    \label{fig:weight}
\end{figure}

\subsection{Quantitative results}
Tuning results for the similarity weight $\alpha$ (in Eq.~\ref{eq:loss}), for all VM methods, are shown in Fig.~\ref{fig:weight} with the best results summarized in Table~\ref{tab:scores}. Note that the DSC is not very sensitive to the choice of the similarity weight for MINE$_{local}$, and the only notable sensitivity for non-positive Jacobian determinants occurs when $\alpha > 5$.  On mono- and multi-modal tasks, all methods achieve comparable results at peak performance with a $\sim20\%$ increase for mono-modal and $\sim14\%$ increase for multi-modal DSC compared to the affine results. (Side-by-side comparison of all methods in terms of dice performance per area are available on request.)

\setlength\tabcolsep{4 pt}
\begin{table}[b]
    \centering
    \begin{threeparttable}[b]
        \caption{Best registration results, per task and method.}
        \begin{tabular}{|l|c|c|c|c|}
            \hline
            \multirow{2}{*}{} & \multicolumn{2}{c|}{Mono-modality}                               & \multicolumn{2}{c|}{Multi-modality}                                 \\ \cline{2-5} 

            Framework   & Mean Dice                 & $|J_{\phi}|\leq0$  
                        & Mean Dice                 & $|J_{\phi}|\leq0$  \\ \hline
            Affine      & $57.6 \:(10.8)$   & -                          
                        & $54.2 \:(11.9)$   & -  \\                 \hline
            ANTs        & $77.7 \:(3.6)$   & $4.0 \:(32.3)$      
                        & $68.7 \:(10.7)$   & $2.1 \:(9.7)$ \\ \hline 
            VM-CC       & $77.3 \:(3.7)$   & $0.3 \:(1.4)$       
                        & $67.7 \:(8.5)$   & $0 \:(0)$ \\    \hline
            VM-MINE$_{g}$ & $76.7 \:(3.6)$   & $0 \:(0)$         
                        & $67.4 \:(8.5)$   & $0 \:(0)$ \\    \hline
            VM-MINE$_{l}$ & $76.7 \:(3.6)$       & $0 \:(0)$         
                        & $68.2 \:(8.2)$   & $0 \:(0)$ \\    \hline
        \end{tabular}
        \label{tab:scores}
    \end{threeparttable}
\end{table}

Even though ANTs combines multiple similarity metrics and SyN, it does not substantially outperform the VM-based methods and has more outliers in failed registrations, making the deep-learning-based approaches more reliable for clinical applications. Comparing the VM-based methods for multi-modal registration in terms of DSC, MINE$_{local}$ performs better than CC, with MINE$_{global}$ a close third place. For mono-modal registration, CC is slightly better than MINE$_{local}$ and MINE$_{global}$. However, there are some limits to how strongly we can extrapolate from these findings, as there can be imperfections in automated segmentation, pre-processing performance across multiple modalities, and observed run-to-run variations in mean DSC differences (of up to $0.5\%$).
Table~\ref{tab:scores} and Fig.~\ref{fig:weight} show that the diffeomorphic properties of VM are unaffected by the switch to multi-modal registration and choice of similarity metric. The number of non-diffeomorphic transforms are negligible in all cases up to the peak-performance point, based on DSC, though they rapidly increase thereafter.

Since the similarity metric does not need to be evaluated at test-time, run-time of VM is the same for all methods at $0.46\:(\pm\:0.02)$ seconds in the mono-modal tasks, increasing to $0.60\:(\pm\:0.02)$ seconds for multi-modal registration on a Titan Xp GPU. In both tasks, deep-learning-based methods are orders of magnitudes faster than ANTs (approx. $10^4\:(\pm\:10^3)$ seconds), and are fast enough for most real-time applications.

The low sensitivity to $\alpha$, together with the similar DSC performance, lower outliers, and faster execution, makes MINE$_{local}$ the overall best, or at least equally favourable to VM-CC, in both mono- and multi-modal tasks on the basis of the quantitative results.

\begin{figure}[b!]
    \centering
    \includegraphics[trim= 7 0 58 25, clip, width=8.5cm]{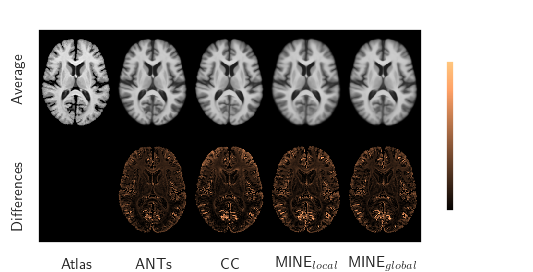}
    \includegraphics[trim= 0 1 0 28, clip, width=8.5cm]{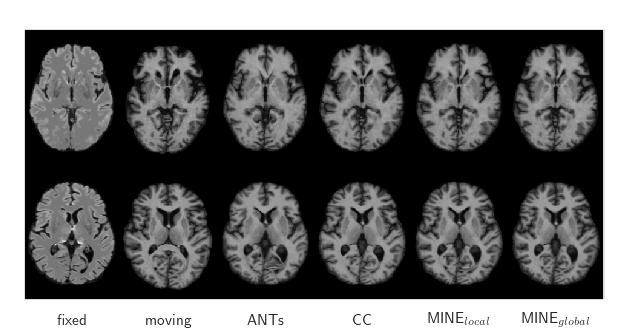}
    \caption{Top: average results for the mono-modal registration after aligning all images to the atlas and absolute difference with the atlas. Bottom: Representative samples of multi-modal registration.}
    \label{fig:averages}
\end{figure}

\subsection{Qualitative results}
Representative registration results are shown in  Fig.~\ref{fig:averages}. All methods produce accurate results on both tasks with minor differences in varying locations. Ignoring the extracortical differences resulting from background masking in the atlas, the most prominent differences for all methods are in the cortex where the largest inter-patient anatomical variation occurs, though differences in the lateral ventricles are also visible. In the mono-modal task these differences are consistent throughout the image for ANTs and MINE while VM-CC shows larger average misalignment in the frontal lobe, and especially the anterior horn of the lateral ventricles, compared to the rest of the image. This is consistent for all values of $\alpha$ in all similarity metrics evaluated. We hypothesize that intensity normalization in small neighborhoods for CC, while producing excellent overall results, can have a negative effect on learning features for registration where there is large anatomical variation. Since MINE is a global metric, even in our local sampling strategy, it does not suffer from this variation. The difference images also show that MINE behaves consistently everywhere with minimal differences on sub-structure boundaries where voxel intensities change. 

\begin{figure}[ht]
    \centering
    \includegraphics[trim=20 0 40 20, clip, width=8.5cm]{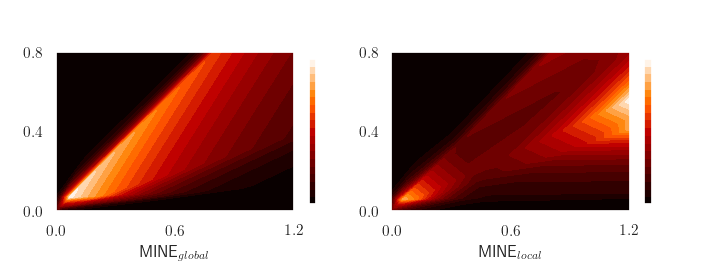}
    \caption{Multi-modal joint distribution responses ($\mathbb{P}$ in Eq.~\ref{eq:mi:lowerbound}) to artificial inputs over the data range.
    Compared to the $MINE_{global}$, the sampling strategy of MINE$_{local}$ results in a more complex response that is a closer approximation of, and guides learning towards, the true joint distribution of an aligned multi-modal image pair.}
    \label{fig:jointMINE}
\end{figure}

To further investigate the MINE-network we took two artificial inputs that cover the range of possible intensity-pair combinations present in the training data to visualize the network's joint-distribution prediction map $\mathbb{P}$, see Fig.~\ref{fig:jointMINE} for multi-modal results. The local shuffling strategy has removed less informative background-brain intensity-pairs compared to MINE$_{global}$, instead using local brain-brain-pairs, as seen in unaligned inputs. This is shown by the more complex joint-distribution for MINE$_{local}$, using a 2-layer CNN, creating better features for registration.

In clinical applications it is important to avoid non-diffeomorphic transformations as these represent cases where anatomically implausible results are generated (e.g. structures disappear or break apart). For real-time applications it is very important to maintain good regional performance (measured by DSC) and anatomical fidelity (measured by Jacobian determinants) and having fast, sub-second run-times. The local form of the MINE-network allows us to satisfy all three of these criteria, and without difficulty in tuning hyper-parameters. This makes MINE-local an excellent candidate for real-time clinical applications, from both a theoretical and practical standpoint.

\section{Conclusion}
We have shown that a small 2-layer CNN MINE-network is capable of producing state-of-the-art mono- and multi-modal registration results on continuous data when smart sampling strategies are used. This eliminates the cumbersome quantization step of conventional MI-based registration, and has relatively low sensitivity to the only key hyper-parameter -- the image similarity training weight. This makes the similarity metric easy to use, which provides accurate results to tune training, with an extremely low rate of producing non-diffeomorphic transformations or failures/outliers. Overall, it produces accurate and anatomically correct results that are fast enough for most real-time applications of multi-modal deformable registration in image-guided therapies.

\bibliographystyle{IEEEbib}
\bibliography{bibliography}

\end{document}